\documentclass[english,keywords,twocolumn,amsmath,amssymb, showpacs]{revtex4-2}
\usepackage[T1]{fontenc}
\usepackage{graphicx}
\usepackage{color}
\usepackage{bm}
\usepackage{longtable}
\usepackage{amsmath}
\usepackage{amsfonts}
\usepackage{amssymb}
\usepackage{subfigure}
\usepackage{bbold}
\usepackage{bbm}
\usepackage{rotating}
\usepackage{verbatim}
\usepackage{braket}
\usepackage{multirow}
\usepackage{hyperref}
\usepackage[english]{babel}
\usepackage{comment}
\usepackage{float}
\usepackage{ulem}
\usepackage{blkarray, bigstrut}
\newcommand{\lorena}[1]{{\color[rgb]{0,0,0}{#1}}}
\begin{document}
\title{Temperature-induced measurement sensitivity enhancement via imaginary weak values}
\author{L. B. Ferraz$^{1}$}
\author{A. Matzkin$^{1}$}
\author{A. K. Pan$^{2}$}
\affiliation{$^{1}$ Laboratoire de Physique Théorique et Modélisation (LPTM) – UMR CNRS 8089, CY Cergy Paris Université, Cergy-Pontoise, France}
\affiliation{$^{2}$ Department of Physics, Indian Institute of Technology Hyderabad, Kandi, Sangareddy, Telengana-502284, India.}

%\footnote{lorena.ballesteros-ferraz@cyu.fr}
%\footnote{akp@nitp.ac.in}
\begin{abstract}
We investigate the potential of weak measurement and post-selection to enhance measurement sensitivity when the initial probe state is mixed. In our framework, the mixedness of the probe's density operator is controlled by temperature. We focus on two key quantities: the signal-to-noise ratio and the quantum Fisher information of the final probe state, evaluated after post-selection is applied on the system. Our analysis employs a rigorous, all-order coupling treatment of measurement, demonstrating that the signal-to-noise ratio can be enhanced in certain scenarios by increasing the temperature. However, this enhancement is fundamentally constrained by the validity conditions of the weak measurement regime. Regarding the quantum Fisher information, we find that for a pure probe state, incorporating post-selection does not improve precision beyond the standard (non-post-selected) strategy when the post-selection probability is accounted for. In contrast, when the initial probe state is mixed, the quantum Fisher information for the probe state after post-selection in the system can surpass that of the standard strategy. Notably, we show that the quantum Fisher information \lorena{might} diverge and grow unboundedly with temperature, illustrating a scenario where thermal noise can, counterintuitively, enhance metrological precision.
\end{abstract}
\maketitle

\section{Introduction}
In quantum mechanics, measurement plays a fundamental role. The precession of a measurement depends not only on technological capabilities, but also on the inherent fundamental constraints imposed by the theory itself. Any improvement of such an `in principle' precision can lead to deeper insights into the underlying nature of the system under investigation. This paper explores the potential utility of weak measurements for improving measurement precision in the presence of temperature-dependent noise.

The pioneering concept of weak measurement in quantum mechanics, originally introduced by Aharonov, Albert, and Vaidman \cite{aav}, reveals that the measured value (called weak value) of an observable can appear anomalous, lying outside the eigenvalue spectrum of the studied observable. Over the years, weak measurement has attracted significant interest due to its ability to reveal seemingly counterintuitive quantum effects, both theoretically \cite{duck,av91,mit,jeff,jozsa, dressel2012significance,brunner,geszti2010postselected, kedem2010modular, zhu2011quantum, zhu2011quantum, koike2011limits, parks2011variance, susa2012optimal, di2012full, dressel12, aha02, lundeenhardy, panmatz, nakamura, wu2011weak, nisi,pan-cheshire,de2022role, cohen2017quantum, svensson2017quantum, ferraz2022geometrical,pan-LG} and experimentally \cite{ritchie,pryde2005measurement, wang2006experimental, hosten,jwvexp,yokota2009direct,sti,steinberg11,lundeen11,zil,kim,str, wang2024high, xu2021specific,exp23}. From a conceptual point of view, weak measurements offer fresh perspectives on foundational issues, such as resolving counterfactual paradoxes \cite{av91,vaidman,aha02,mit}, reconstructing average quantum trajectories of photons \cite{steinberg11}, and challenging macrorealism by violating the Leggett-Garg inequality \cite{will}. Beyond its foundational implications, weak measurements also have practical applications, including the detection of the tiny spin Hall effect of light \cite{hosten}, measurement of minute transverse beam deflections \cite{sti}, enhancement of signal-to-noise ratio in phase estimation via interferometry \cite{str}, and quantum state protection \cite{kim}.

Before proceeding further, let us briefly recapitulate the essential framework of weak measurement \cite{svensson2013pedagogical,wvrev}. Consider a quantum system initially prepared in the pure state $|i\rangle$ (commonly referred to as the pre-selected state), while the associated pointer is initialized in the state $|\psi\rangle=\int dx \psi(x)|x\rangle$. In general, the ancilla does not need to reside in a continuous Hilbert space; it can, for instance, be a discrete system such as a qubit \cite{brun2008test}. Moreover, the initial state does not need to be pure; it may also be a mixed state \cite{dressel2010contextual}. The weak interaction is introduced via a perturbative Hamiltonian of the form 
\begin{equation}
\hat{H}_{I}= \theta(t) \hat A \otimes\hat {P},
\end{equation}
 where $\theta(t)$ is a smooth time-dependent coupling function satisfying  $\int_{0}^{t}\theta(t)dt= \theta$, with $t$ denoting the duration of the interaction.

The total system-pointer state after the interaction can be expressed as
\begin{equation}
|\Psi^{\prime}\rangle = \int dx e^{-i \theta \hat{A}\otimes \hat{P}}\left(\, \psi(x)\, |x\rangle\otimes |i\rangle\right).
\end{equation}

The crucial next step involves post-selection: a strong projective measurement is performed on the system, selecting a specific final state $|f\rangle$. As a result, the pointer is left in the state
\begin{equation}
|\psi_f\rangle = \langle f | \Psi^{\prime} \rangle = \int dx\langle f | e^{-i \theta \hat{A}\otimes \hat{P}}\left(\, \psi(x)\, |x\rangle\,  \otimes|i\rangle\right).
\end{equation}
Assuming that the coupling strength $\theta$ is small, indicating a weak interaction, we can expand the exponential and neglect higher-order terms in $\theta$. Under this weak interaction approximation, the final pointer state becomes
\begin{equation}
|\psi_f\rangle \approx \langle f | i \rangle\int e^{-i \theta A_w \hat{P}}dx\, \psi(x)\, |x\rangle\, , 
\end{equation}
where 
\begin{align}
	A_{w}=\frac{\langle {f}| \hat A|{i}\rangle}{\langle {f}|{i}\rangle}
\end{align}
is the weak value of the observable $\hat A$ and $|\langle {f}|{i}\rangle|^{2}$ is the post-selection probability.

% \lorena{By appropriately choosing the final state $|f\rangle$ and the initial state $|i\rangle$, one can achieve a small post-selection probability, allowing the weak value $A_w$ to lie far outside the eigenvalue spectrum of the corresponding observable. Interestingly, $A_w$ can also be complex~\cite{jozsa}, yet this does not imply any physical inconsistency. The imaginary part of the complex weak value results in a shift of the pointer in the space representation where the perturbation is introduced, while the real part causes a shift in the conjugate space. Specifically, in the above case, the change in the expectation value of the pointer position is proportional to the real part of the weak value, $\langle \hat{X} \rangle_f - \langle \hat{X} \rangle_i \propto \Re(A_w)$,
%and the change in the momentum expectation is proportional to its imaginary part,$\langle \hat{P}_x \rangle_f - \langle \hat{P}_x \rangle_i \propto \Im(A_w)$, where $\langle \hat{X} \rangle_f$ denotes the expectation value of the pointer observable in the post-selected state, and similarly for the others.}

A large weak value leads to a substantial shift in the post-selected meter distribution, often beyond the range of eigenvalue shifts, resulting in an amplification of the detector signal. This amplification effect can be particularly useful for detecting tiny physical effects~\cite{hosten}. The imaginary part of the weak value has found important applications in quantum metrology. In recent years, the use of weak measurements to enhance measurement sensitivity has attracted significant attention. It has been claimed that phase estimation using an imaginary weak value can outperform standard interferometric techniques~\cite{brunner}. Furthermore, studies have shown that the precision of the measurements can be improved even in the presence of technical noise~\cite{sti,kedem}. In particular, Kedem~\cite{kedem} demonstrated that the signal-to-noise ratio can be enhanced through post-selection.

The estimation of a phase using white light has also been demonstrated experimentally~\cite{xu}. It is important to highlight two closely related, yet distinct concepts: signal amplification and the efficient estimation of a small parameter. While weak measurements can produce amplified signals, often at the cost of a low post-selection probability, such amplification does not necessarily imply a metrological advantage. In other words, an enhanced detector signal alone is not sufficient to guarantee improved precision in parameter estimation.

Interestingly, Hofmann pointed out the connection between weak values and the Cramér–Rao bound for estimating an unknown parameter~\cite{hofmann, hofmann2012estimation}. The fundamental limits of amplification~\cite{pang, dil1} and strategies for its optimization have also been discussed. More recently, a renewed discussion has been initiated by Tanaka and Yamamoto~\cite{tanaka}, who investigated whether post-selection and weak measurements can offer any metrological advantage, specifically whether they can increase quantum Fisher information in parameter estimation compared to standard strategies. Their analysis is based on a pure probe state. However, the general consensus is negative: several studies~\cite{knee, fer} have shown through different lines of reasoning that post-selection does not enhance the Fisher information. Furthermore, it has been demonstrated~\cite{knee} that even in the presence of technical noise and for finite ensembles, the Fisher information cannot be increased through post-selection.

In this paper, we provide a concrete example to rigorously examine two key metrological quantities relevant to parameter estimation: the signal-to-noise ratio and the quantum Fisher information. Our analysis is conducted in the presence of technical noise affecting the initial preparation of the probe. Specifically, we consider temperature-dependent noise and demonstrate that, under such conditions, the signal-to-noise ratio can be enhanced with increasing temperature. This result is consistent with the findings of Kedem~\cite{kedem}. A careful analysis reveals that the signal-to-noise ratio does not increase indefinitely; instead, it reaches a maximum at a specific temperature determined by the system parameters. 

Furthermore, we demonstrate that, for a pure probe state, the quantum Fisher information of a suitably chosen post-selected sub-ensemble can never surpass that of the full ensemble used in a standard measurement strategy without post-selection. However, the scenario changes when the probe state is mixed due to technical noise in the initial preparation. We show that the imaginary part of the weak value plays a crucial role in this case. Specifically, we demonstrate that the quantum Fisher information can be enhanced by the presence of technical noise, mediated through the imaginary weak value. Throughout this paper, we consider a Gaussian probe state and assume the technical noise to be white, with zero mean and finite variance. We specifically consider the probe state as a Gaussian subjected to a Maxwell-Boltzmann momentum distribution. Since such a distribution is controlled by temperature, the degree of mixedness increases with temperature. We demonstrate that, in the weak coupling regime, the signal-to-noise ratio improves as the mixedness of the probe state increases. Additionally, we identify the optimal degree of mixedness for the meter state and the optimal choices of pre- and post-selected states.

The structure of this paper is as follows. In Section~\ref{section:presentation_of_the_system}, we introduce the studied system, in Section~\ref{section:signal_to_noise_ratio}, we present a comprehensive analysis of the signal-to-noise ratio enhancement for the temperature-dependent mixed state in comparison to the pure state, focusing on specific scenarios that meet the validity conditions for weak measurements. Section~\ref{section:fisher_information} explores the quantum Fisher information for both pure and mixed states, highlighting the temperature-dependent increase in Fisher information. Finally, in Section~\ref{section:conclusions}, we summarize the key findings and conclude the article.
\section{Temperature dependent mixed probe state}\label{section:presentation_of_the_system}
We consider the total initial state of the system and the meter to be described by the density operator $\hat{\rho}$. For simplicity, we assume that $\hat{\rho}$ is a product state of the system and the meter, such that $\hat{\rho} = \hat{\rho}_s \otimes \hat{\rho}_d$,
where $\hat{\rho}_s$ and $\hat{\rho}_d$ denote the density operators of the system and the meter, respectively. In this work, we consider the system to be initially in a pure state, $\hat{\rho}_s = |i\rangle \langle i|$, while the probe (meter) state $\rho_d$ is taken to be mixed. The mixedness of the probe is controlled by its temperature. To model this, we consider the initial meter wave function to be
\begin{equation}
\label{wavefn1}
\langle x|\psi \rangle=(2 \pi  \sigma^2)^{-1/4} e^{-x^2/4\sigma^2 + i p_0 x\lorena{/\hbar}},
\end{equation} 
where $\sigma$ is the initial width of the associated wave packet.

The initial momentum is given by $p_0=m v_0$, where $v_0$ is the group velocity of the wave packet propagating along the $+\widehat{\bf x}$ axis. However, if the preparation is not ideal, meaning the system cannot be initialized in a perfectly identical state across all trials, the particles described by Eq.~\ref{wavefn1} may instead be in thermal equilibrium, characterized by a thermal distribution of initial velocities. In this case, each particle still has a wave function of the form Eq.~\ref{wavefn1}, but with its initial velocity drawn from a Maxwell-Boltzmann distribution, given by
\begin{equation}
n(p_0)=(2\pi  m k_{B} T)^{-1/2} e^{-p_0^2/2m k_{B} T}\lorena{,}
\end{equation}
where $m$ is the mass of the particle, $k_{B}$ is the Boltzman constant and $T$ denotes the temperature of the particles.

The initial thermal state of the ensemble can be described by the following density operator
\begin{equation}
\label{eq:initial_thermal_state}
    \hat{\rho}_d = |\psi\rangle\langle\psi| = \int dx \int dx^{\prime} \int dp_0\, n(p_0)\, \psi(x)\, \psi^*(x')\, |x\rangle \langle x'|,
\end{equation}
where $n(p_0)$ is the thermal distribution of initial momenta. The elements of this density operator in position representation are given by
\begin{eqnarray}
\label{mixedstate}
\langle x|\hat{\rho}_{d}|x^{\prime}\rangle=(2 \pi  \sigma^2)^{-1/2}e^{-\frac{x^2+(x^{\prime})^2}{4 \sigma^2} + \frac{m k_{B} T (x-x^{\prime})^2}{2\lorena{\hbar^2}}}.
\end{eqnarray}
Alternatively, in the momentum representation, the elements of the density operator are given by:

\begin{eqnarray}
&&\langle p|\hat{\rho}_{d}|p^{\prime}\rangle=\frac{\sqrt{2 \sigma^2}}{\sqrt{\pi (\hbar^2+ 2\alpha )}}\\
\nonumber 
&\times& exp\left[-\frac{\hbar^2\sigma^2\left(p^2+p'^2\right)+2k_bm\sigma^4T\left(p-p'\right)^2}{\hbar^4+4\hbar^2k_bm\sigma^2T}\right],
\end{eqnarray}
where $\alpha=2 m k_{B}T\sigma^2$.

Following the standard approach, the purity of the density operator can be quantified by evaluating the trace of its square, $\mathrm{Tr}[\rho_d^2]$, given by
\begin{align}
	Tr[\hat{\rho}_{d}^2]=\frac{\hbar}{\sqrt{\left(\hbar^2+ 4 m k_{B}T \sigma^2\right)}},
\end{align}
indicating that the degree of mixedness of the density operator is strongly dependent on the temperature $T$.

Clearly, at $T = 0$, we have $\mathrm{Tr}[\hat\rho_d^2] = 1$, indicating that $\hat\rho_d$ represents a pure state. However, for any $ T > 0 $, $\mathrm{Tr}[\hat\rho_d^2] < 1$, reflecting an increase in mixedness. As the temperature increases, the state becomes progressively more mixed, and for a fixed $\sigma$, the density matrix $\hat{\rho}_d$ can approach a maximally mixed state at sufficiently high temperature.
\section{Temperature induced SNR improvement}\label{section:signal_to_noise_ratio}
We begin by introducing the concept of signal-to-noise ratio (SNR) within the context of a weak measurement with post-selection.
\subsection{Signal-to-noise Ratio}

%\lorena{I am very much against acronyms. It really makes the reading difficult and it is a real problem for people with dyslexia. For me, we can remove all acronyms of the paper.}

In an interferometric setup, the signal is quantified by the average shift in the pointer's position or momentum variables, given by
\begin{equation}
    \delta x = \left| \langle \hat{X} \rangle_f - \langle \hat{X} \rangle_i \right| \quad \text{\&} \quad \delta p = \left| \langle \hat{P} \rangle_f - \langle \hat{P} \rangle_i \right|,
\end{equation}
where no post-selection of a specific system state is performed. Here, $\langle \hat{X} \rangle_i$ and $\langle \hat{X} \rangle_f$ denote the expectation values of the position observable in the initial and final states, respectively. The corresponding noise in this process is characterized by the square root of the variance of the relevant variables, given by
\begin{equation}
    \Delta x = \sqrt{\langle \hat{X}^2 \rangle_f - \langle \hat{X} \rangle_f^2} \quad \text{\&} \quad \Delta p = \sqrt{\langle \hat{P}^2 \rangle_f - \langle \hat{P} \rangle_f^2}.
\end{equation}

The signal-to-noise ratio (SNR) for a single measurement is given by $\delta x / \Delta x$ (or $\delta p / \Delta p$), where $x$ ($p$) denotes the pointer variable being observed. After $N$ independent trials, the variance is reduced by a factor of $\sqrt{N}$, leading to an improved SNR of
\begin{eqnarray}
\label{snrweak}
\mathcal{S}_x=\frac{\sqrt{N}\delta x}{\Delta x}; \mathcal{S}_p=\frac{\sqrt{N}\delta p}{\Delta p}.
\end{eqnarray}

If the interaction is very weak, the resulting average shifts in the pointer coordinates will be small compared to the intrinsic noise of the measurement apparatus, leading to a low signal-to-noise ratio. However, by employing a weak measurement scheme, it is possible to enhance the SNR. If $N$ trials are performed and the probability of successful post-selection is $P$, then the number of useful (i.e., successfully post-selected) trials is $N P$. Consequently, the signal-to-noise ratio in the post-selected scenario must be redefined as
\begin{eqnarray}
\mathcal{S}^{\prime}_x = \frac{\delta x _{ps}\sqrt{NP}}{\Delta x _{ps}}; \mathcal{S}^{\prime}_p = \frac{\delta p _{ps}\sqrt{NP}}{\Delta p _{ps}}
\end{eqnarray}
where $\delta x_{ps}= |\langle \hat{X}\rangle_{ps} - \langle \hat{X}\rangle_i|$,  $\delta p_{ps}= |\langle \hat{P}\rangle_{ps} - \langle \hat{P}\rangle_i|$, and $\langle\hat{P}\rangle_{ps}=Tr[\rho_{d}^{ps}\hat{P}]$, $\langle\hat{X}\rangle_{ps}=Tr[\rho_{d}^{ps}\hat{X}]$. Similarly for $\Delta x _{ps}$ and $\Delta p _{ps}$.

\subsection{SNR for any arbitrary coupling strength}
We now consider an arbitrary coupling strength, followed by post-selection of the system in a specific state, represented by the projection operator $\hat\Pi_{f} = |f\rangle\langle f|$, which, in principle, requires an additional strong measurement.

To model this, we consider the interaction Hamiltonian $\hat H = g(t)\, \hat{A} \otimes \hat{P}$, where $\hat{A}$ is the relevant observable of the system, and $g(t)$ is a smooth function of time satisfying $\int_{0}^{t} g(t) dt = \theta$ that determines the coupling strength. Here, $t$ denotes the duration of the interaction. The total density operator after time evolution is then given by
\begin{align}
	\hat\rho^{\prime}=e^{-i \theta \hat A\otimes \hat P}\rho \ e^{i \theta \hat A\otimes \hat P}\lorena{,}
\end{align}
which can be represented in terms of sinusoidal functions as, by considering observables that are their own inverses, i.e., $\hat A^2 = I$,
\begin{eqnarray}
\hat{\rho}'&=&\left(\hat{I}\otimes \cos{\left(\theta\hat{P}\right)}-i\hat{A}\otimes\sin{\left(\theta\hat{P}\right)}\right)\hat{\rho} \\ \nonumber
&\times&\left(\hat{I}\otimes \cos{\left(\theta\hat{P}\right)}+i\hat{A}\otimes\sin{\left(\theta\hat{P}\right)}\right).
\end{eqnarray}
%which can be written by using Cauchy summation formula as
%\begin{align}
%	\rho^{\prime}=\sum_{n=0}^{\infty} \frac{1}{n!}\left[ \sum_{k=0}^{n} (-1)^{k}\left(\begin{array}{c} 
%n\\ k\end{array}\right)\left\{(A\otimes P)^{k}\rho \ (A\otimes P)^{n-k}\right\} \right]
%\end{align} 
If the system is subsequently post-selected in the state $\hat\Pi_{f} \otimes \hat I$, the resulting meter state is
\begin{eqnarray}
\label{finalprobe}
	\hat\rho^{ps}_{d}=\frac{Tr_{s}[(\hat\Pi_{f}\otimes \hat I)\hat\rho^{\prime}]}{Tr[(\hat\Pi_{f}\otimes \hat I)\hat\rho^{\prime}]}\lorena{.}
\end{eqnarray}
Then, the numerator of Eq.~\ref{finalprobe} is
\begin{eqnarray}
&&Tr_{s}[(\hat\Pi_{f}\otimes \hat I)\hat\rho^{\prime}]=Tr_{s}[\hat\Pi_{f}\hat\rho_{s}]\\
\nonumber
&\times&(\cos{\left(\theta\hat{P}\right)}\hat{\rho}_d\cos{\left(\theta\hat{P}\right)}-iA_w\sin{\left(\theta\hat{P}\right)}\hat{\rho}_d\cos{\left(\theta\hat{P}\right)}\\ \nonumber
&+&iA_w^{*}\cos{\left(\theta\hat{P}\right)}\hat{\rho}_d\sin{\left(\theta\hat{P}\right)}+|A_w|^2 \sin{\left(\theta\hat{P}\right)}\hat{\rho}_d\sin{\left(\theta\hat{P}\right)}),
\end{eqnarray}
and the denominator is
\begin{eqnarray}
Tr[(\hat\Pi_{f}\otimes \hat I)\hat\rho^{\prime}]=Tr_s[\hat\Pi_{f}\hat\rho_{s}]\times \mathcal{Z}_{1},
\end{eqnarray}
with

\begin{eqnarray}
\mathcal{Z}_{1}&=&Tr[(\cos{\left(\theta\hat{P}\right)}\hat{\rho}_d\cos{\left(\theta\hat{P}\right)}-iA_w\sin{\left(\theta\hat{P}\right)}\hat{\rho}_d\cos{\left(\theta\hat{P}\right)}\\ \nonumber
&+&iA_w^{*}\cos{\left(\theta\hat{P}\right)}\hat{\rho}_d\sin{\left(\theta\hat{P}\right)}+|A_w|^2 \sin{\left(\theta\hat{P}\right)}\hat{\rho}_d\sin{\left(\theta\hat{P}\right)})], 
\end{eqnarray}
where 
\begin{eqnarray}
A_w=\frac{Tr[\hat\Pi_{f}\hat A\hat\rho_{s}]}{Tr[\hat\Pi_{f}\hat\rho_{s}]};  
A_w^{\ast}=\frac{Tr[\hat\Pi_{f}\hat\rho_{s} \hat A]}{Tr[\hat\Pi_{f}\hat\rho_{s}]}
\end{eqnarray}
and

\[|A_w|^{2}=\frac{Tr[\hat\Pi_{f}\hat A\hat\rho_{s}\hat A]}{Tr[\hat\Pi_{f}\hat\rho_{s}]}.\]
Note that the weak measurement condition has not yet been applied. For the mixed probe state given by Eq.~\ref{mixedstate}, and retaining all orders of the coupling strength, the signal-to-noise ratio takes the form of
\begin{widetext}
\begin{eqnarray}
\label{eq:SNR_mixed}
&&(S^{\prime}_{p})^{\text{mixed}}=\frac{\sqrt{N}|\langle f|i\rangle|\sqrt{\beta}\theta\text{Im}\left(A_w\right)}{\sqrt{\left(\hbar^2\theta^2-\sigma^2+2\alpha\theta^2\right)\epsilon+e^{\theta^2\omega'}\sigma^2\left(1+|A_w|^2\right)\left(\epsilon+e^{\theta^2\omega'}\left(1+|A_w|^2\right)\right)-4\left(\hbar^2+2\alpha\right)\theta^2\text{Im}\left(A_w\right)}},
\end{eqnarray}
\end{widetext}
where $\beta=\hbar^2+2\alpha$, $\epsilon=|A_w|^2-1$, $\omega'=\frac{\hbar^2+2\alpha}{2\sigma^2}$. Since $\langle \hat{P} \rangle_{i} = 0$, the pointer shift is given by $\delta p_{ps} = \mathrm{Tr}[\rho_{d}^{ps} \hat{P}]$. The signal-to-noise ratio for an initial ancilla pure state, obtained by setting $T = 0$, is denoted by $(\mathcal{S}_{p}^{\prime})^{\text{pure}}$, and takes the following form
\begin{widetext}
\begin{eqnarray}
    (S^{\prime}_{p})^{\text{pure}}=\frac{2\hbar\theta\sqrt{N}|\langle f|i\rangle||\text{Im}\left(A_w\right)|}{\sqrt{\left(\left(-\sigma^2+\hbar^2\theta^2\right)\left(-1+|A_w|^2\right)+e^{\frac{\hbar^2\theta^2}{2\sigma^2}}\sigma^2\left(1+|A_w|^2\right)\right)\left(1+e^{\frac{\hbar^2\theta^2}{2\sigma^2}}\left(1+|A_w|^2\right)-|A_w|^2\right)-4\hbar^4\theta^2|\text{Im}\left(A_w\right)|^2}}.
\end{eqnarray}
\end{widetext}
\subsection{Improvement of SNR with temperature}
Note again that the quantities $(\mathcal{S}_{p}^{\prime})^{\text{pure}}$ and $(\mathcal{S}_{p}^{\prime})^{\text{mixed}}$ are derived without invoking the weak coupling condition.
 For a pre-selected state $|i\rangle = \cos\phi\,|0\rangle + i\sin\phi\,|1\rangle$, a post-selected state $|f\rangle = |0\rangle$, and the observable $\hat{A} = \hat{\sigma}_x$, the corresponding weak value is given by $A_w = i\tan\phi$. Using an imaginary weak value, it can be shown that the ratio $\mathcal{A} = (\mathcal{S}_{p}^{\prime})^{\text{mixed}} / (\mathcal{S}_{p}^{\prime})^{\text{pure}}$ can be enhanced for given values of the relevant parameters $\sigma$, $g$, and $\Im(A_w)$. For example, by taking $k_b=3.167\cdot10^{-6}$ in atomic units, $\hbar=1$, $m=50$ atomic units, $T=0$ K, $A_w = 2.31i$, $\theta=0.025$, $N=1000$, $|\langle f|i\rangle|^2=0.001$, $\sigma=1$, is $(\mathcal{S}_{p}^{\prime})^{\text{pure}} = 0.058$. Using the same parameter values, we obtain $(\mathcal{S}_{p}^{\prime})^{\text{mixed}} = 0.61$ for $T = 100$ K. The ratio $\mathcal{A}$ can be very large, indicating a significant improvement in SNR due to the presence of noise in the initial preparation for a given value of $\sigma$. Thus, with all other parameters fixed, a certain form of metrological advantage can be achieved using the mixed state, where temperature plays a key role in controlling the mixedness.

\begin{figure}[htp]
    \centering
    \includegraphics[width=7cm]{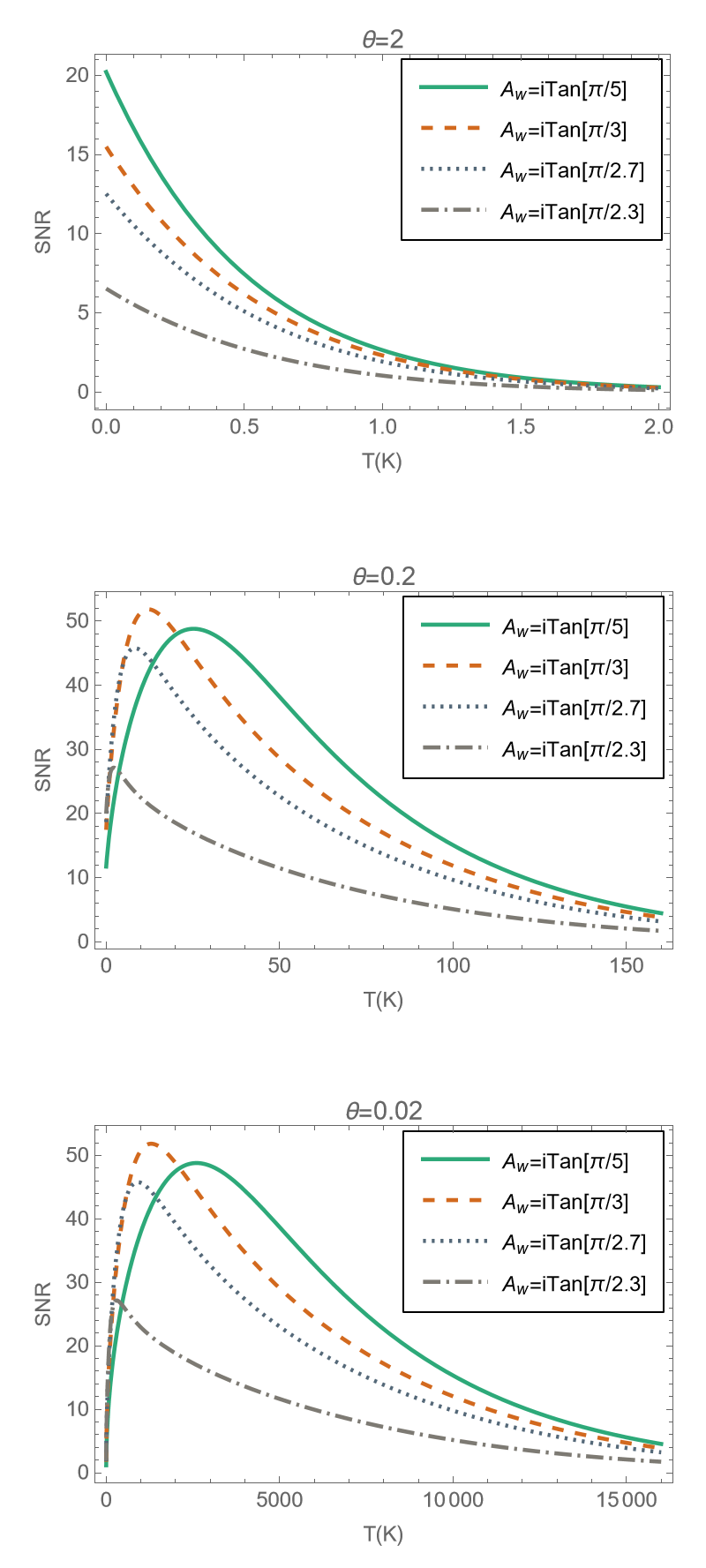}
    \caption{The signal-to-noise ratio (Eq.~\ref{eq:SNR_mixed}) is plotted as a function of temperature for four different weak values and three distinct coupling constants $\theta$. The parameters used are $\sigma = 1$, $\hbar = 1$, $k_B = 3.167 \times 10^{-6}$, $m = 50$ atomic units, $N = 1 \times 10^{4}$, with $|\langle f | i \rangle|^2 = \cos^2 \phi$ and $A_w = i \tan \phi$. \lorena{In these plots, since the interaction strength is not always weak, the post-selection probability is evaluated as $P = \text{Tr}\left[(\hat\Pi_{f}\otimes \hat I)\hat\rho^{\prime}\right]$.}
}
\label{fig:SNR_three_cases}
\end{figure}

We now consider the weak coupling condition, which intuitively corresponds to retaining only the first-order terms in the coupling constant. Under this approximation, by neglecting higher-order terms in $\theta$ in Eq.~\ref{eq:SNR_mixed}, we obtain
\begin{eqnarray}
(S^{\prime}_{p})_{w}^{mixed}=\sqrt{N} |\langle f|i\rangle| \theta\ \Im(A_{w})\sqrt{2\omega'}\lorena{.}
\end{eqnarray}
In this context, the condition for the validity of the weak measurement approximation is given by $\theta\, \Im(A_{w}) \sqrt{2\omega} \ll 1$. Substituting the expression for $\omega$, the explicit form of $S^{\prime}_{p}$ can be written as
\begin{eqnarray}
\label{snrw}
(S^{\prime}_{p})_{w}^{mixed}=\sqrt{N} |\langle f|i\rangle| \frac{ \theta\ \Im(A_{w})\sqrt{1+ 4 m K_{B}T\sigma^2}}{\sigma}. \\ \nonumber
\end{eqnarray}
\lorena{Setting $T = 0$ yields the signal-to-noise ratio for the pure probe state. From Eq.~\ref{snrw}, we observe that for a fixed large value of $\sigma$, the SNR can be enhanced by increasing the temperature $T$, provided the condition for the validity of the weak measurement approximation is satisfied. It is important to note that weak measurements do not necessarily yield the maximum SNR in all scenarios; rather, the post-selection process plays a central role. These results are consistent with the findings of Kedem~\cite{kedem}, as they pertain to the same regime of interest.}
\lorena{As shown in Fig.~\ref{fig:SNR_three_cases}, the behavior of the signal-to-noise ratio as a function of temperature varies significantly with the coupling constant, $\theta$. For weak measurements (small $\theta$), the SNR initially increases with temperature, reaching a maximum at a certain point before decreasing. When the coupling strength is moderate (not too weak or too strong) the SNR begins at a value notably above zero, increases to a peak, and then decreases as temperature continues to rise. In contrast, for strong coupling, the SNR curve lacks a peak altogether: it starts at a specific value and monotonically decreases as temperature increases. In general, the stronger the coupling, the more rapidly the SNR decays toward zero.

This variation in behavior arises from the structure of the ancilla's density operator after measurement. At zero temperature ($T=0$), strong coupling leads to the emergence of multiple peaks in the diagonal elements of the density operator. For example, at $\theta=2$, two peaks are observed; as $\theta$ increases further, more peaks appear. On the other hand, for weak coupling, only a single peak exists at $T=0$. As the temperature increases, a second peak emerges, coinciding with the rise in SNR. However, after two peaks have formed, raising the temperature further not only broadens these peaks but also introduces new ones, which reduces the contrast between states and lowers the SNR.

This suggests that having two well-separated peaks in the ancilla's density matrix is important for effectively measuring a two-level observable such as $\hat{\sigma}_z$. Beyond this point, higher temperatures degrade the measurement quality by reducing the distinguishability of the peaks.}
%\begin{figure}[h]
%{\resizebox{9.0cm}{18.0cm}{\includegraphics{SNR_various_theta_22.png}}}
%\label{fig:SNR_three_cases}\caption{}
%\end{figure}

\section{Temperature induced improvement of the accuracy of the phase estimation}\label{section:fisher_information}
We now investigate the metrological advantage induced by temperature in weak, intermediate, and strong measurement regimes in the presence of post-selection. A central goal in quantum metrology is to determine how efficiently small parameters can be estimated.

The accuracy of such an estimation is fundamentally limited by the Cramér–Rao bound, which sets the lowest achievable variance for unbiased estimators, regardless of the specific observable being measured. In the fundamental scenario where a small parameter $\theta$ is encoded through unitary dynamics $\hat U=e^{-i \theta \hat A}$ and estimated via output probabilities, the Cramér–Rao bound takes the form:
\begin{align}
\delta \theta\geq \frac{1}{\sqrt{N I_{F}(\hat \rho_{\theta})}}
\end{align}
where $\hat \rho_{\theta}$ denotes the parameter-dependent output state of the ancilla after the full measurement protocol, and $I_{F}(\hat \rho_{\theta})$ represents the corresponding Fisher information.

Helstrom \cite{helstorm} provided a natural definition of the quantum Fisher information using the symmetric logarithmic derivative $L_{\theta}$, such that
\begin{equation}
\label{eq:Fisher_information_general}
    I_{F}(\rho_\theta)=Tr[\rho_{\theta}L_{\theta}^2],
\end{equation}
where the symmetric logarithmic derivative satisfies 
\begin{equation}
\label{eq:logarithmic_derivative}
    \frac{\partial\hat{\rho}_{\theta}}{\partial\theta}=\frac{\hat{L}_{\theta}\hat{\rho}_{\theta}+\hat{\rho}_{\theta}\hat{L}_{\theta}}{2}.
\end{equation}
In the case of a pure state, where $\hat\rho_{\theta}=|\psi_{\theta}\rangle\langle\psi_{\theta}|$, the quantum Fisher information simplifies to the form 
\begin{equation}
    I_{F}(\rho_{\theta})=4\left(\langle\psi_{\theta}'|\psi_{\theta}'\rangle -|\langle\psi_{\theta}'|\psi_{\theta}\rangle|^2\right),
\end{equation} 
where a prime ($'$) indicates a derivative with respect to $\theta$. However, for mixed states, it is generally more difficult to obtain a closed-form expression for the quantum Fisher information \cite{toth}. Note that an increase in Fisher information $I_{F}(\rho_{\theta})$ corresponds to an improvement in estimation accuracy. Our goal here is to investigate whether post-selection and weak measurement enhance the Fisher information $I_{F}(\hat \rho^{ps}_{d})$ of the post-selected final probe state $(\hat \rho_{d}^{ps})$, compared to a standard strategy without post-selection.

To investigate both the weak and strong measurement regimes, we employ two distinct strategies. In the weak regime, analytical calculations are feasible. In contrast, the strong regime requires numerical methods. We begin by presenting the results for the weak regime, followed by those for the strong interacting case.

\subsection{Quantum Fisher information in the weak regime}

In the weak regime, since our probe state is Gaussian, we adopt a standard procedure to calculate the Fisher information for Gaussian states \cite{pinel}. Specifically, we aim to evaluate the Fisher information $I_{F}(\hat\rho^{ps}_{d})$ for the post-selected meter density matrix $\hat\rho_{d}^{ps}$, as defined in Eq.~\ref{finalprobe}.

The Wigner function corresponding to the post-selected state $\hat\rho_d^{ps}$ is defined as
\begin{eqnarray}
\label{eq:general_Wigner_funciton}
W(x,p)=\frac{1}{2\pi\hbar}\int_{-\infty}^{\infty} dq^{\prime} e^{-i  x q^{\prime}}\langle p-q^{\prime}|\hat\rho_{d}^{ps}|p+q^{\prime}\rangle.
\end{eqnarray}
The density matrix $\hat\rho_{d}^{ps}$ depends explicitly on the parameter $\theta$, which we aim to estimate. Assuming $\theta$ is small, as appropriate in the weak measurement regime, we expand Eq.~\ref{finalprobe} in powers of $\theta$ and retain terms up to the first order. Following this expansion, we compute the Wigner function of the final ancilla state using Eq.~\ref{eq:general_Wigner_funciton}.
To calculate the quantum Fisher information, we first recall the well- known form of the Wigner function for a $\theta$-dependent Gaussian state, which can be written as
\begin{eqnarray}
\label{wig2}
W({\bf X})_{\theta}=\frac{1}{2\pi \sqrt{|\Sigma_{\theta}|}} e^{-\frac{1}{2} \left({\bf X}-\overline{{\bf X}}_{\theta}\right)\Sigma_{\theta}^{-1}\left({\bf X}-\overline{{\bf X}}_{\theta}\right)},
\end{eqnarray}
where $\overline{{\bf X}}_{\theta}$ denotes the vector of expectation values of the field quadratures in the final ancilla state, $\overline{{\bf X}}_{\theta,i}=Tr[(\hat\rho^{ps}_{d}) X_i]$, while $\bf X$ represents the vector of quadrature variables. The number of quadrature components depends on the number of modes (or particles) in the system. For a single-mode ancilla, there are two quadratures: position ($x$) and momentum ($p$), ${\bf X}=(x,p)^{T}$. The quantity  $|\Sigma_{\theta}|$  denotes the determinant of the covariance matrix $\Sigma_{\theta}$, whose elements are given by
\begin{align}
	\Sigma_{\theta,ij}=\frac{1}{2}\langle X_{i}X_{j}+X_{j}X_{i}\rangle-\langle X_{i}\rangle\langle X_{j}\rangle,
\end{align}
where each $X_{a}$ represents a field quadrature. The ordering of the quadratures in the covariance matrix must match the ordering in the vector $\bf X$ to ensure consistency.

A key tool for deriving the quantum Fisher information is the Bures distance between two density matrices, as introduced in~\cite{caves}. The Bures distance is defined as
\begin{align}
D_{\text{Bures}}(\hat\rho_{1},\hat\rho_{2})=\sqrt{2-2\sqrt{F(\hat\rho_{1},\hat\rho_{2})}}\lorena{,}
\end{align}
where $F(\hat\rho_{1},\hat\rho_{2})=\left[Tr(\sqrt{\hat\rho_{1}}\hat\rho_{2}\sqrt{\hat\rho_{1}})^{1/2}\right]^2$  denotes the Uhlmann fidelity between the two density matrices $\hat\rho_{1}$ and $\hat\rho_{2}$. For our purpose, the Fisher information can be derived from the Bures distance between two neighboring density matrices $\hat\rho_{\theta}$ and $\hat\rho_{\theta+\epsilon}$ using the relation
\begin{align}
\label{fisdefn}
	I_{F}(\hat\rho_{\theta})=4\left[\left(\frac{\partial D_{Bures}(\hat\rho_{\theta},\hat\rho_{\theta+\epsilon})}{\partial\epsilon}\right)|_{\epsilon=0}\right]^2
\end{align}
For the Gaussian state defined in Eq.~\ref{wig2}, the quantum Fisher information can be calculated using Eq.~\ref{fisdefn} as
\begin{align}
\label{eq:Fisher_general_Gaussian}
	I_{F}(\hat\rho_{\theta})=\frac{Tr[(\Sigma_{\theta}^{-1}\Sigma_{\theta}^{\prime})^2]}{2(1+P_{\theta}^2)}+\frac{2 (P_{\theta}^{\prime})^2}{1-P_{\theta}^{4}}+(\Delta{\bf X}_{\theta}^{\prime})^{T}\Sigma_{\theta}^{-1}\Delta{\bf X}_{\theta}^{\prime},
\end{align}
where the prime ($'$) denotes differentiation with respect to $\theta$; $\Delta{\bf X}_{\theta}^{\prime}=d\langle{\bf X}_{\theta+\epsilon}-{\bf X}_{\theta}\rangle/d\epsilon|_{\epsilon=0}$, and $P_{\theta}$ quantifies the purity of the state, defined as
\begin{equation}
\label{eq:formula_for_P_theta}
    P_{\theta}=Tr[(\hat\rho_{d}^{ps})^2]=\frac{1}{\sqrt{(det\Sigma_{\theta})}}.
\end{equation}
We begin by calculating the quantum Fisher information under the assumption that the meter state is initially pure and no post-selection is applied. In this case, the final state after the interaction is given by
\begin{equation}
\ket{\psi'} = \int dx e^{-i\theta \hbar \hat{A} \otimes \hat{P}} \ket{i} \ket{x} \psi(x),
\end{equation}
where $\psi\left(x\right)$ denotes the initial meter wavefunction, which is obtained by setting $T=0$ in the density operator described in Eq.~\ref{eq:initial_thermal_state}. The wavefunction is given by
\begin{equation}
\label{eq:wavefunction_Fisher}
\psi\left(x\right)=\left(\frac{2\sigma^2}{\pi\hbar^2}\right)^{1/4}e^{-\frac{\sigma^2x^2}{\hbar^2}}.
\end{equation}

The quantum Fisher information for the standard measurement strategy, without invoking post-selection or assuming the weak measurement regime, is given by
\begin{align}
	I_{F}^{pure}(\hat\rho^{\prime})=\frac{\hbar^2\langle\hat{A}^2\rangle_i}{\sigma^2}=\frac{\hbar^2}{\sigma^2},
\end{align}
where $\langle\hat{A}^2\rangle_i$ corresponds to the expectation value of the operator $\hat{A}^2$ in the initial system state $\ket{i}$.

When post-selection is applied, the interaction is weak, and the initial ancilla state is pure, the quantum Fisher information is
\begin{align}
	I_{F}^{pure}(\rho^{ps}_{d})=\frac{\lorena{\hbar^2}\left|A_{w}\right|^2}{\sigma^2},
\end{align}
which is independent of the parameter $\theta$, in agreement with the result obtained by Tanaka and Yamamoto\cite{tanaka}.

The ratio $\mathcal{R}^{pure}=I_{F}^{pure}(\rho^{ps}_{d})/I_{F}^{pure}(\rho^{\prime})$ compares the quantum Fisher information with and without post-selection, assuming a pure ancilla state. This ratio simplifies to $\mathcal{R}^{pure}=\left|A_{w}\right|^2$, which can exceed unity for suitable choices of pre- and post-selection states. This might suggest that post-selection enhances the quantum Fisher information. However, if the post-selection probability is also taken into account, the effective ratio becomes $\mathcal{R}^{pure}=|\langle f|A|i\rangle|^{2}$, which is bounded by the eigenvalues of the operator $\hat{A}$. Hence, no advantage appears. 

We now consider the case where the probe state is mixed, as described by Eq.~\ref{finalprobe}, with its mixedness controlled by the temperature of the preparation process. Under the weak measurement condition, retaining terms up to first order in the coupling parameter $\theta$, the explicit form of the Wigner function is
\begin{align}
\label{wmixed}
	W_{mixed}=\frac{1}{4\pi \sqrt{\beta}}e^{-\frac{x^2}{8\sigma^2}-\frac{2 p^2 \sigma^2}{\beta}+4\theta p \Im[A_w]+\frac{\hbar\theta x \Re[A_w]}{\sigma^2}}   
\end{align}
where $\beta=\hbar^2+2\alpha$, and $\alpha=2mk_BT\sigma^2$. By comparing Eq.~\ref{wmixed} with the general form in Eq.~\ref{wig2}, we observe that the expression obtained from our measurement protocol in the weak regime corresponds to a Gaussian state. Appendix~\ref{appendix:calculation_fisher_weak} provides a detailed breakdown of the components involved in the calculation of the quantum Fisher information. The final expression for the quantum Fisher information in the weak regime is given by
\begin{align}
\label{eq:Fisher_information_weak_regime}
	I_{F}^{mixed}(\rho^{ps}_{d})=\frac{\lorena{\hbar^2}\left|A_{w}\right|^2 + 2\alpha (\Im(A_w))^2}{\sigma^2},
\end{align}
where the dependence on $T$ is linear due to the term proportional to $\alpha$. \lorena{Without post-selection, the Fisher information takes the form $I_{F}^{\text{pure}}(\rho^{\prime}) = \frac{\hbar^{2}}{\sigma^{2}}$
for a pure initial state. In contrast, for a mixed state at low temperature it becomes $I_{F}^{\text{mixed}}(\rho^{\prime}) = \frac{\hbar^{2}\langle \hat{A} \rangle_{i}^{2}}{\sigma^{2}}$, and decreases to zero as the temperature increases. Since $\hat{A}^{2}=\hat{A}$ ensures $\langle \hat{A} \rangle_{i} \leq 1$, the use of a mixed initial state necessarily lowers the achievable precision in estimating $\theta$. Therefore, raising the initial ancilla temperature does not provide any advantage in precision.} However, when post-selection is considered in the weak regime, the ratio of Fisher information between the cases of a pure and a mixed initial ancilla state is $\mathcal{R}^{post}=I_{F}^{mixed}(\rho^{ps}_{d})/I_{F}^{pure}(\rho^{ps}_{d})=1+\frac{2\alpha\left(Im\left(A_w\right)\right)^2}{\hbar^2 |Aw|^2}$. Taking into account the post-selection probability and substituting $\alpha = 2 m K_B T \sigma^2$, this ratio becomes the form given by
\begin{align}
	\mathcal{R}^{post}=1+ \frac{4 m K_{B} T (\Im(\langle f|A|i\rangle))^2\sigma^2}{\hbar^2|A_w|^2}|\braket{f|i}|^2.
\end{align}
This leads us to conclude that the Fisher information increases as the temperature rises.

\lorena{\subsection{Quantum Fisher information at arbitrary interaction strength}}
\lorena{Next, we aim to compute the quantum Fisher information beyond the weak measurement regime. In these cases, analytical calculations become extremely involved. Therefore, we adopted a numerical approach to evaluate the quantum Fisher information.

To do so, we use the general expression for the quantum Fisher information given in Eq.~\ref{eq:Fisher_information_general}, where the symmetric logarithmic derivative operator can be obtained from Eq.~\ref{eq:logarithmic_derivative}, provided the density operator $\hat{\rho}_{\theta}$ is known. Assuming that the density operator $\hat{\rho}_{\theta}$ can be diagonalized with eigenvalues $d_{\alpha}$, the symmetric logarithmic derivative operator in the diagonal basis of the density operator, denoted $\hat{L}_{\theta}^{d}$, can be computed using the following expression:
\begin{equation}
\label{eq:logarithmic_derivative_diagonal}
    \hat{L}_{\theta}^{d} = 2 \sum_{i,j} \frac{\left(\frac{\partial \hat{\rho}_{\theta}}{\partial \theta}\right)^{d}_{ij}}{d_i + d_j} \ket{i} \bra{j},
\end{equation}
where $\hat{L}_{\theta}^{d} = \hat{P}^{-1} \hat{L}_{\theta} \hat{P}$ and $\left(\frac{\partial \hat{\rho}_{\theta}}{\partial \theta}\right)^{d}_{ij} = \left(\hat{P}^{-1} \frac{\partial \hat{\rho}_{\theta}}{\partial \theta} \hat{P}\right)_{ij}$. Here, $\hat{P}$ is the matrix whose columns are the eigenvectors of the density operator $\hat{\rho}_{\theta}$.}
\begin{figure}[h!]
{{\includegraphics[scale=0.40]{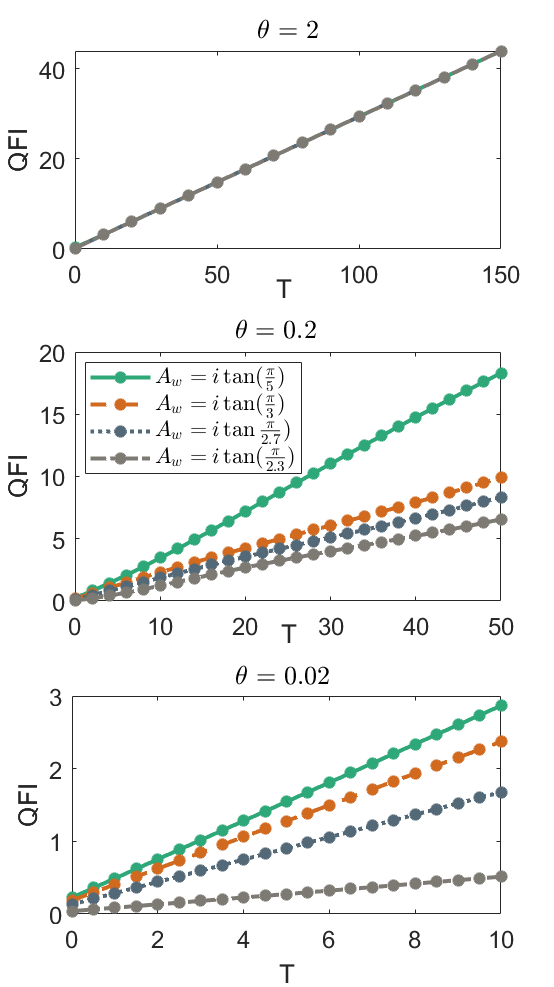}}}
\caption{\label{fig:QFI_middle_strong}\lorena{\footnotesize Quantum Fisher information (QFI) as a function of the temperature of the initial ancilla’s mixed state, taking into account the post-selection probability. The results are shown for three different values of interaction strength $\theta$ and four different weak values, $A_w$. The quantum Fisher information increases with the imaginary part of the weak value only in the weak measurement regime, provided the post-selection probability is included. The parameters employed are $\sigma = 1$, $\hbar = 1$, $k_B = 3.167 \times 10^{-6}$, $m = 50$ atomic units, with $|\langle f | i \rangle|^2 = \cos^2 \phi$ and $A_w = i \tan \phi$. In these plots, since the interaction strength is not always weak, the post-selection probability is evaluated as $P = \text{Tr}\left[(\hat\Pi_{f}\otimes \hat I)\hat\rho^{\prime}\right]$.}}
\end{figure}

\lorena{To diagonalize the density operator $\hat{\rho}_{\theta}$, it is first necessary to discretize it. In our analysis, we work in the momentum representation. Once the operator is discretized over an appropriate momentum range, the quantum Fisher information can be computed using the expression given in Eq.~\ref{eq:Fisher_information_general}. The results of these calculations are presented in Fig.~\ref{fig:QFI_middle_strong}.

As illustrated in the plots, the quantum Fisher information generally increases with temperature. For sufficiently high temperatures, this increase is linear. However, in the lower temperature range, the behavior of the quantum Fisher information depends on the interaction strength, $\theta$. At both large interaction strength and at small $\theta$ within a low-temperature range, a linear trend is also observed. For small interaction strengths, the increase in quantum Fisher information slows at a certain temperature, reflecting a change in trend. A larger imaginary part of the weak value does not lead to an enhancement in quantum Fisher information. For large values of $\theta$ ($\theta = 2$), the quantum Fisher information becomes independent of the weak value once the post-selection probability is taken into account.
%The distinct behavior in the weak regime is likely attributed to the emergence of a second peak in the diagonal of the density operator at a specific temperature. In comparison, for intermediate and strong interaction strengths, the density operator exhibits two peaks already at zero temperature. 

To understand the behavior of the quantum Fisher information in the limit of infinite temperatures, we use
an analytical model, detailed in Appendix~\ref{appendix:calculation_limit_T_infinite_fisher_weak} that indicates a diverging behavior as the temperature approaches infinity, provided that the post-selection probability remains nonzero. The model approximates the ancilla distribution by a uniform distribution between symmetric finite values of a maximal momentum component. By letting this maximal momentum go to infinity, we show in Appendix~\ref{appendix:calculation_limit_T_infinite_fisher_weak} that the Fisher information becomes unbounded, and this would ideally enable the estimation of parameters with arbitrarily high precision.

Note that this phenomenon is not unique and has also been observed in other continuous-variable systems~\cite{zhang2012unbounded, pinel}, where it is remarked that this behavior may be seen as an artifact of requesting
unphysical ressources as the relevant quantities go to infinity. 

\section{conclusions}\label{section:conclusions}
We considered two key metrological quantities, the signal-to-noise ratio and the quantum Fisher information to evaluate the potential advantage of post-selection in weak measurement scenarios. Importantly, we extended our analysis beyond the weak regime to include all interaction strengths, from weak to strong coupling.

Our findings indicate that, in a specific scenario within the weak measurement regime, temperature-dependent noise can enhance measurement sensitivity in terms of an increased signal-to-noise ratio when post-selection is applied. However, this enhancement remains bounded and cannot surpass the shot-noise limit.

In agreement with recent studies~\cite{kedem,knee}, we confirm that post-selection does not increase the quantum Fisher information for a pure probe state when the post-selection probability is taken into account. Nevertheless, when the probe is in a mixed state, where the degree of mixedness is controlled by the temperature of the initial preparation, the quantum Fisher information can increase significantly, indicating improved parameter estimation accuracy. Remarkably, the quantum Fisher information appears to diverge in the limit of infinite temperature.

When employing the mixedness of the ancilla to enhance the measurement protocol, it is essential to consider both metrological quantities simultaneously. While the quantum Fisher information benefits from high temperatures, the signal-to-noise ratio requires more careful tuning to achieve optimal performance. Since the signal-to-noise ratio can rapidly decrease beyond its optimal point, the most effective strategy for enhancing measurement precision is to maximize the signal-to-noise ratio while maintaining the temperature as high as possible without surpassing the optimal threshold.}
\\

{\it Acknowledgements:} Author acknowledge the support from DST/ ICPS/QuEST/Theme-1/4. This research was supported by the EUTOPIA Science and Innovation Fellowship Programme and funded by the European Union Horizon 2020 programme under the Marie Skłodowska-Curie grant agreement No 945380.

%\clearpage
\onecolumngrid
\vspace{3cm}
\appendix
\lorena{\section{Quantum Fisher Information in the weak measurement regime}
\label{appendix:calculation_fisher_weak}
In this appendix, we provide the detailed expressions for the components of Eq.~\ref{eq:Fisher_general_Gaussian} required to compute the Fisher information in the weak regime, using the Wigner function given in Eq.~\ref{wmixed}.

To compute the $2 \times 2$ covariance matrix $\Sigma_{\theta}$, we evaluate the following four terms:
\begin{eqnarray}
    \Sigma_{\theta, xx}&=&\langle \hat{X}^2\rangle-\langle\hat{X}\rangle^2=\int dx\int dp x^2 W_{\text{mixed}} - \left(\int dx\int dp x W_{\text{mixed}}\right)^2 \approx 4 \sigma^2 \\ \nonumber
    \Sigma_{\theta, xp}&=&\Sigma_{\theta, px}=\frac{1}{2}\langle \hat{X}\hat{P}+\hat{P}\hat{X}\rangle-\langle\hat{X}\rangle\langle\hat{P}\rangle=\int dx\int dp xp W_{\text{mixed}} - \left(\int dx\int dp x W_{\text{mixed}}\right)\left(\int dx\int dp p W_{\text{mixed}}\right)\approx 0 \\ \ \nonumber
    \Sigma_{\theta, pp}&=&\langle \hat{P}^2\rangle-\langle\hat{P}\rangle^2=\int dx\int dp p^2 W_{\text{mixed}} - \left(\int dx\int dp p W_{\text{mixed}}\right)^2 \approx \frac{\hbar^2}{4\sigma^2}+k_BmT \\ \nonumber.
\end{eqnarray}
As can be seen, the matrix $\Sigma_{\theta}$ does not depend on $\theta$ to first order. Consequently, its derivative with respect to $\theta$ is zero. The quantity $P_{\theta}$ can be computed from the determinant of $\Sigma_{\theta}$, as given in Eq.~\ref{eq:formula_for_P_theta}. The result is
\begin{equation}
    P_{\theta}=\frac{1}{\sqrt{\hbar^2+4k_bmT\sigma^2}},
\end{equation}
which is independent of $\theta$, and therefore its derivative with respect to $\theta$ is also zero.

Finally, we calculate $\Delta{\bf X}_{\theta}^{\prime}=d\langle{\bf X}_{\theta+\epsilon}-{\bf X}_{\theta}\rangle/d\epsilon|_{\epsilon=0}$. To do this, we proceed as follows: first, we obtain the Wigner function at $\theta+\epsilon$ by substituting $\theta$ with $\theta+\epsilon$. Next, we compute the expectation values of both position and momentum for the difference of Wigner functions $W_{\text{mixed}, \theta+\epsilon}-W_{\text{mixed}, \theta}$. Finally, we differentiate with respect to $\epsilon$ and evaluate the result at $\epsilon=0$. The elements of the vector are then
\begin{eqnarray}
    \Delta X'_{\theta}&=&\frac{d \left[\int dx\int dp x\left(W_{\text{mixed}, \theta+\epsilon}-W_{\text{mixed}, \theta}\right)\right]}{d\epsilon}|_\epsilon=2\hbar Re\left(A_w\right) \\ \nonumber
    \Delta P'_{\theta}&=&\frac{d \left[\int dx\int dp p\left(W_{\text{mixed}, \theta+\epsilon}-W_{\text{mixed}, \theta}\right)\right]}{d\epsilon}|_\epsilon=\frac{Im\left(A_w\right)\left(\hbar^2+4k_bmT\sigma^2\right)}{2\sigma^2}.
\end{eqnarray}
Using these elements, the quantum Fisher information in the weak regime can be calculated, and the result is given by Eq.~\ref{eq:Fisher_information_weak_regime}.}

\lorena{\section{High-temperature limit of quantum Fisher information}
\label{appendix:calculation_limit_T_infinite_fisher_weak}
In this appendix, we calculate the limiting form of the quantum Fisher information as the temperature of the initial system state, given in Eq.~\ref{eq:initial_thermal_state}, approaches infinity. To do this, we assume that in this limit, we can take the initial ancilla state to be described by the identity operator,
\begin{equation}
    \hat{\rho}_a=\int dx \ket{x}\bra{x}=\int dp \ket{p}\bra{p}.
\end{equation}
We will actually consider a symmetric bounded version of this equation in momentum space, first assuming that only momenta
in the range $\{-p_{max},p_{max}\}$ contribute, before taking $p_{max}$ to infinity at the last step. In this case, the ancilla state after the measurement protocol, including post-selection, is given by
\begin{equation}
\label{eq:ancilla_state_infinite_temperature}
    \hat{\rho}_d^{ps}=\frac{1}{N}\left[\cos{\left(\theta\hat{P}\right)}\hat{\rho}_a\cos{\left(\theta\hat{P}\right)}-iA_w\sin{\left(\theta\hat{P}\right)}\hat{\rho}_a\cos{\left(\theta\hat{P}\right)}+iA_w^{*}\cos{\left(\theta\hat{P}\right)}\hat{\rho}_a\sin{\left(\theta\hat{P}\right)} +|A_w|^2\sin{\left(\theta\hat{P}\right)}\hat{\rho}_a\sin{\left(\theta\hat{P}\right)}\right], 
\end{equation}
where $\hat{\rho}_a = \hat{I}$ and $N$ is a normalization constant defined as
\begin{equation}
    N=\int_{-p_{max}}^{p_{max}}dp\left(\cos^2{\left(\theta p\right)}+2\text{Im}\left(A_w\right)\sin{\left(\theta p\right)}\cos{\left(\theta p\right)}+|A_w|^2\sin^2{\left(\theta p\right)}\right).
\end{equation}
Since the ancilla state after the measurement protocol Eq.~\ref{eq:ancilla_state_infinite_temperature} is diagonal, its derivative is also diagonal. Therefore, the diagonal elements of the operator $\hat{L}_{\theta}$ can be written as
\begin{equation}
    L_{\theta}^d=\frac{\left(\frac{\partial\rho_{\theta}}{\partial\theta}\right)^d}{\rho_{\theta}^d}.
\end{equation}
The Fisher information $I_F$ is therefore given by
\begin{equation}
    I_F=\int_{-\text{p}_{\text{max}}}^{\text{p}_{\text{max}}} dp \hat{\rho}_d^{ps}\left(p,p\right) \hat{L}_{\theta}^2\left(p,p\right). 
\end{equation}
We now compute the leading-order behavior of the Fisher information, $I_F$, in the limit $\text{p}_{\text{max}} \to \infty$. In this regime, the leading term is given by $\frac{4\text{p}_{\text{max}}^2}{3}$ which diverges as $\text{p}_{\text{max}} \to \infty$. Therefore within this model the Fisher information becomes unbounded in the high-temperature limit.}
\twocolumngrid
\bibliographystyle{unsrt}
\bibliography{biblio}
\end{document}